\documentclass[sigconf]{acmart}
\AtBeginDocument{%
  }

\copyrightyear{2026}
\acmYear{2026}
\setcopyright{cc}
\setcctype{by}
\acmConference[IDC '26]{Proceedings of the 25th Interaction Design and Children Conference}{June 22--25, 2026}{Brighton, United Kingdom}
\acmBooktitle{Proceedings of the 25th Interaction Design and Children Conference (IDC '26), June 22--25, 2026, Brighton, United Kingdom}
\acmDOI{10.1145/3773077.3806118}
\acmISBN{979-8-4007-2283-7/2026/06}

\usepackage{multirow}
\usepackage{longtable}
\usepackage{float}
\usepackage{appendix}
\usepackage{enumitem}
\usepackage{booktabs}

\newif\ifcameraready

\begin{document}

\title{Engagement Is Not Transfer: A Withdrawal Study of a Consumer Social Robot with Autistic Children at Home}

\author{Yibo Meng}
\authornote{Both authors contributed equally to this research.}
\affiliation{%
  \institution{Cornell University}
  \city{Ithaca}
  \country{United State}
}
\email{yim4007@med.cornell.edu}

\author{Guangrui Fan}
\authornotemark[1]
\affiliation{%
  \institution{Taiyuan University of Science and Technology}
  \city{Taiyuan}
  \country{China}
}
\email{962557032@qq.com}

\author{Bingyi Liu}
\affiliation{%
  \institution{University of Michigan, Ann Arbor}
  \city{Ann Arbor}
  \state{Michigan}
  \country{United State}
}
\email{bingyi@umich.edu}

\author{Yingfangzhong Sun}
\affiliation{%
  \institution{Politecnico di Milano}
  \city{Milan}
  \country{Italy}
}
\email{yingfangzhong.sun@mail.polimi.it}

\author{Ruiqi Chen}
\affiliation{%
  \institution{University of Washington}
  \city{Seattle}
  \state{Washington}
  \country{United States}
}
\email{ruiqich@uw.edu}

\author{Haipeng Mi}
\affiliation{%
  \institution{Tsinghua University}
  \city{Beijing}
    \country{China}
}
\email{mhp@tsinghua.edu.cn}

\renewcommand{\shortauthors}{Meng et al.}

\begin{abstract}
This study examines whether engagement with social robots translates into improved human-directed social abilities in autistic children. We conducted an 8-week home-based randomized controlled trial with 40 children aged 5–9 using a commercial social robot (Qrobot). Families were assigned to either continued robot access or robot withdrawal. Quantitative measures and caregiver interviews assessed anxiety, social motivation, emotion inference, and empathy. Results showed that continued robot access significantly reduced anxiety, confirming strong affective benefits and high usability. However, children in the withdrawal group demonstrated greater improvements in social motivation, emotion understanding, and empathic behaviors toward caregivers and peers. Qualitative findings revealed a “handoff versus siloing” pattern: withdrawal promoted reorientation toward human social interaction, while continued access concentrated engagement within the child–robot dyad and limited transfer to real-world contexts. We interpret these results as evidence that high engagement does not guarantee social transfer.
\end{abstract}



\begin{teaserfigure}
  \includegraphics[width=\textwidth]{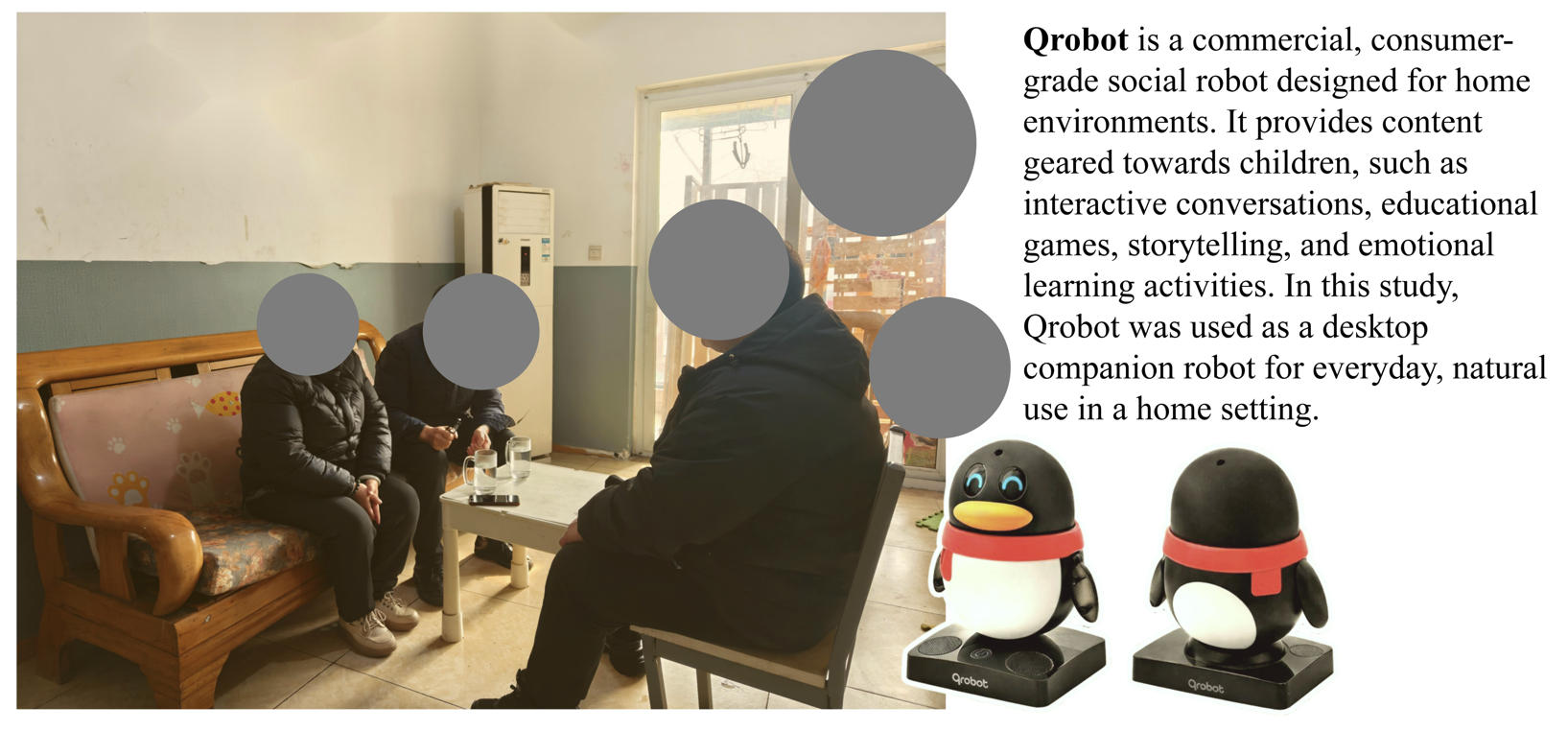}
  \caption{The qualitative interview process involving the children's guardians and researchers, along with an introduction to Qrobot.}
  \label{fig0}
\end{teaserfigure}
\maketitle
\section{Introduction}
Social robots are increasingly explored as supports for autistic children across clinical, educational, and home contexts. Recent syntheses suggest that robot-assisted interventions can reliably elicit interactional behaviors and yield benefits on some targeted outcomes, but also highlight substantial heterogeneity in study designs, outcome measures, and ecological validity---making it difficult to infer what children gain beyond the immediate human--robot interaction (HRI) moment \cite{kouroupa2022meta,kohli2023robot,vagnetti2024clinical}. At the same time, the field is moving ``from lab to reality,'' with growing emphasis on effectiveness under real-world constraints (e.g., family routines, caregiver mediation, sustained use, and uneven adherence) \cite{david2025_lab_to_reality}. This shift is especially relevant for consumer-grade systems that enter homes as everyday companions rather than as tightly scripted clinical tools \cite{gomezespinosa2024early,santos2023applications}. In such contexts, the central question is no longer only whether a robot can engage a child with autism, but what this engagement does to the child's broader social ecology over time.

A recurring assumption in HRI for autism is that a robot functions as a \emph{social scaffold}: a simplified, predictable partner that helps children practice interaction skills that later generalize to human--human interaction (HHI) \cite{ghiglino2023artificial}. This scaffolding view motivates work that targets higher-order social cognition (e.g., theory of mind) and aims to translate robot-guided learning into downstream human social functioning \cite{ghiglino2025tom}. However, evidence for durable generalization is mixed, and follow-up work underscores that post-intervention trajectories depend on the broader interaction context, caregiver practices, and how the system is integrated into daily life \cite{ghiglino2021followup}. In parallel, emerging research has begun to explicitly design robots as \emph{conversational catalysts} intended to enhance long-term HHI at home, shifting the focus from child--robot dyads to child--robot--family ecologies \cite{chen2025catalysts,meng2026mistyforest}. Taken together, these developments point to a critical evaluation gap: high engagement and smooth interaction can coexist with weak transfer if learned behaviors remain bounded within the robot context.

As robots become more available and more ``companion-like,'' a plausible failure mode is that the robot concentrates interaction within the child--robot dyad---raising \emph{substitution-risk} concerns, i.e., robot-centered engagement that does not translate into human-directed bids---rather than functioning as a bridge to HHI. Importantly, strict interaction-time displacement is difficult to establish without usage logs or time-allocation measures; in this paper we treat displacement as a hypothesis and focus on transfer-relevant outcomes and caregiver accounts. These concerns are echoed across stakeholder perspectives, ethical scholarship, and longitudinal accounts: professionals worry about replacement dynamics in autism care \cite{ashwini2024looks}; ethical frameworks call for evaluation extending to long-term socio-emotional effects and dependency risks \cite{langer2023ethical}; and longitudinal accounts of ``retired'' robots confirm that children may continue relating to robots as meaningful social artifacts well after formal use ends \cite{zhao2025robotstayed}. Together, this motivates a central dilemma for interaction design: systems optimized for comfort and engagement may inadvertently reduce incentives to practice the uncertainty and reciprocity of human social life.

One of the most consistent findings across robot-assisted interventions is improvement in affective regulation, particularly anxiety reduction \cite{wu2025anxiety}. Yet recent work testing the social motivation theory of autism suggests that co-occurring anxiety can meaningfully shape social behavior, complicating simple causal stories in which reduced anxiety automatically yields increased social seeking \cite{bagg2024socialmotivation}. This creates a design paradox: the very properties that make robots effective anxiolytic companions (predictability, low-demand reciprocity, constant availability) may also make them \emph{too} attractive as social partners, potentially reducing motivation to engage with humans.

A core limitation of many robot studies is that they primarily measure performance or engagement \emph{in the presence of the robot}. If the key promise is scaffolding, the most diagnostic evidence should appear when the robot is \emph{absent}: do social bids and emotion understanding persist and reappear in HHI contexts, or do they remain confined to the child--robot dyad? We argue that \emph{withdrawal} is not merely a control condition; it can operate as a \emph{transfer probe} that reveals whether a robot functioned as a bridge to HHI or as a self-contained social destination. We note, however, that withdrawal may reorganize the home ecology through multiple pathways---including caregiver co-activity, substitutions to other media, and shifts in family routines---so observed behavioral reallocations should be interpreted as reflecting the broader interaction ecosystem rather than the robot's absence alone. This perspective complements recent calls for more ecologically grounded evaluation of robot-assisted therapy \cite{david2025_lab_to_reality} and aligns with design efforts that explicitly aim to increase HHI rather than only robot-directed engagement \cite{chen2025catalysts}.

We conducted an 8-week, home-based, parallel-group randomized controlled trial (RCT) to examine the ``engagement is not transfer'' dilemma in a consumer-grade social robot system (Qrobot). Children with autism who had prior experience using the system were randomly assigned either to (1) \emph{continued access} to the robot or (2) a \emph{withdrawal} condition in which the robot was removed from the home. Using a mixed-methods design, we tracked trajectories across three time points (baseline, midpoint, post) using standardized measures of social motivation, emotion understanding, empathy-related behavior, and anxiety, and complemented these with semi-structured guardian interviews. We additionally assessed perceived usability of the HRI to determine whether observed outcomes could plausibly be attributed to interaction friction or poor interface design.

Guided by the scaffolding versus substitution risk dilemma, we ask:
\begin{itemize}
    \item \textbf{RQ1:} Can highly engaging and usable social robot interaction coexist with weaker improvements in \emph{human-directed} social motivation and emotion inference (relative to withdrawal) in home deployments?
    \item \textbf{RQ2:} What does robot \emph{withdrawal} reveal about transfer---specifically, how do children's social bids and emotion-attunement behaviors reallocate when the robot is absent?
    \item \textbf{RQ3:} What interaction design implications follow from divergent trajectories in engagement, anxiety regulation, and human-directed social development?
\end{itemize}

This paper makes four contributions to interaction design and child-centered HRI. \emph{Empirically}, we report mixed-method evidence of divergent trajectories in a home deployment: continued access was associated with improved affect regulation (lower anxiety), while human-oriented outcomes improved more under withdrawal. \emph{Conceptually}, we operationalize the scaffolding-to-substitution risk as an evaluative target, distinguishing engagement/comfort outcomes from transfer outcomes. \emph{Methodologically}, we introduce withdrawal as a \emph{transfer probe} for child--robot interventions, showing how studying the robot's absence can reveal ecological reallocation dynamics that are invisible in robot-present evaluations. \emph{For design}, we derive implications for ``transfer-first'' social robots---including handoff mechanisms and planned off-ramps intended to preserve anxiolytic benefits without reinforcing robot-centered social containment.

\section{Related Work and Conceptual Framework}
This paper is motivated by a growing tension in child-facing Human--Robot Interaction (HRI): systems that are highly engaging and usable can still fail to support---and may even inadvertently hinder---\emph{transfer} to human--human interaction (HHI). To position our study and define the analytic lens we use throughout the paper, we synthesize recent work (2020--2026) across four threads: (1) how social-robot interventions for autism are commonly evaluated and what ``success'' tends to mean; (2) the persistent challenge of generalization and the central role of caregivers and family ecologies; (3) the robot's frequently observed affective and anxiolytic value, and why that value may complicate transfer; and (4) ethical and ecological concerns specific to deploying social agents in children's homes. We conclude with a conceptual framework---\emph{Engagement is Not Transfer}---that operationalizes the scaffolding-versus-substitution-risk dilemma and motivates the use of withdrawal as a transfer probe.

\subsection{What the Field Often Optimizes: Engagement, Acceptability, and Short-Horizon Outcomes}
Over the past decade, social robots have been increasingly explored as therapeutic or educational supports for autistic children, with the past five years producing multiple systematic reviews of robot-assisted interventions and randomized controlled trials (RCTs) \cite{kouroupa2022_socialrobots_meta,salimi2021_rcts_review,alabdulkareem2022_robot_assisted_therapy_review}. Across these reviews, a recurring pattern is that studies frequently foreground \emph{feasibility} and \emph{acceptability}---e.g., whether children will engage with the robot, tolerate sessions, and complete tasks---often coupled with short-horizon improvements measured within or immediately after the intervention window \cite{kouroupa2022_socialrobots_meta,alabdulkareem2022_robot_assisted_therapy_review}. This emphasis is understandable: engagement is a prerequisite for any intervention with children, and robots may be uniquely capable of sustaining attention through predictable feedback, gamified tasks, and novelty.

However, an engagement-centered evaluation norm can also create an interpretive pitfall. When engagement, enjoyment, or usability scores are treated as proxies for therapeutic value, it becomes easy to overestimate downstream developmental effects and to under-measure unintended consequences in the child's broader social ecology. Recent scholarship has explicitly called for clearer distinctions between efficacy in controlled settings and effectiveness in everyday contexts, where competing demands, family routines, and social opportunities shape what an intervention actually changes \cite{david2025_lab_to_reality}. In parallel, work on long-term HRI highlights that sustained interaction introduces dynamics (habituation, dependency, shifting expectations, evolving household practices) that short studies are structurally unable to capture \cite{matheus2025_long_term_interactions_review}. Taken together, this literature suggests that ``the robot works'' cannot be concluded solely from high acceptability or short-term performance; the field needs evaluation approaches that directly test whether benefits \emph{persist} and \emph{generalize} beyond the robot-mediated context.

\subsection{What Children Need: Transfer, Generalization, and Caregiver-Mediated Ecologies}
If the goal of robot-assisted autism interventions is to improve real-world functioning, then the central scientific and design question is not merely whether children can perform skills \emph{with} the robot, but whether those skills and motivations generalize to HHI. Systematic reviews consistently note heterogeneity in outcomes and methodological variation that complicate conclusions about generalization \cite{kouroupa2022_socialrobots_meta,salimi2021_rcts_review}. This is particularly salient for social competencies: many robot activities simplify the social world by offering consistent turn-taking rules, constrained emotional expressions, and forgiving interaction repair. These properties can reduce cognitive and affective load, but may also reduce exposure to the variability and reciprocity that characterize human interaction.

A growing set of studies therefore reposition caregivers not as external observers but as integral components of the intervention system. Recent reviews of parental involvement in robot-mediated interventions argue that caregiver participation can be feasible and may shape outcomes through co-regulation, modeling, and the creation of opportunities for practice beyond the robot session \cite{piccolo2024_parental_involvement_review}. Empirical work similarly suggests that parental involvement in robot-assisted autism therapy can meaningfully influence intervention effects, highlighting that ``robot + child'' is rarely the true unit of analysis in home settings \cite{amirova2023_parental_involvement_ra_at,zeng2025parental}. In other words, transfer is not only a child-level learning problem; it is also an \emph{ecological} problem that depends on how caregivers orchestrate activities, how routines change, and how social opportunities are redistributed across family members and peers.

Comparative studies that explicitly contrast robot-based and human-based interventions (or examine conditions with different degrees of mediation) further reinforce the importance of designing for generalization rather than assuming it \cite{so2023_robot_vs_human_joint_attention,cao2020_robot_assisted_joint_attention}. These comparisons suggest that the evaluation of robot-assisted interventions should prioritize outcomes that are inherently human-oriented (e.g., spontaneous initiations toward caregivers/peers, sensitivity to real interpersonal affect cues) and should articulate the mechanisms by which robot interaction is expected to translate into HHI.

\subsection{The Robot as an Affective Regulator: Comfort Benefits and the ``Predictability Trade-off''}
A second recurring thread in recent HRI research is the robot's value as an affective regulator: predictable, structured interactions can reduce stress, improve emotional stability, or support calm behavioral states \cite{rakhymbayeva2021_long_term_engagement,matheus2022_ommie_deep_breathing}. This finding is not limited to autism contexts; anxiety-reduction work with social robots demonstrates that guided, predictable routines (e.g., paced breathing) can measurably shift state anxiety and self-reported affect \cite{matheus2022_ommie_deep_breathing,zhao2025immersive,meng2026whalesong}. In autism interventions specifically, long-term engagement studies often describe how stability and predictability can help children sustain participation and reduce the likelihood of overload \cite{rakhymbayeva2021_long_term_engagement}.

Yet the very properties that make robots emotionally ``safe'' may also complicate transfer. A predictable partner can lower anxiety, but may also reduce the child's incentive to practice the more effortful task of interpreting and responding to the ambiguity of human social cues. This sets up a \emph{predictability trade-off}: designs that maximize comfort and minimize uncertainty may inadvertently minimize opportunities to tolerate uncertainty---a capability that is central to real-world social functioning. Recent calls to move from lab efficacy to real-world effectiveness emphasize that such trade-offs are likely to become visible only when robots are embedded in daily routines over longer time spans \cite{david2025_lab_to_reality,matheus2025_long_term_interactions_review}. From an IDC perspective, the implication is that affect regulation benefits should not be evaluated in isolation; they must be interpreted alongside changes in the child's broader social engagement patterns.

\subsection{Ethical and Ecological Stakes of In-Home Social Agents for Children}
Deploying social robots in children's homes raises ethical questions that extend beyond typical usability or safety considerations. Conceptual work on child--robot relationships highlights that companionship, friendship metaphors, and moral development concerns may be implicated when children form durable social bonds with artificial agents \cite{constantinescu2022_children_robot_friendship}. At the same time, child-centered AI governance guidance emphasizes that systems interacting with children should prioritize wellbeing, privacy, transparency, and accountability, particularly when deployment contexts reduce opportunities for professional oversight \cite{unicef2025_ai_for_children_policy_guidance}. These concerns are amplified in autism contexts, where children may be especially sensitive to routines and may experience interventions as emotionally meaningful.

For our purposes, the ethical lens is not only about risk prevention; it also clarifies \emph{what should count as success}. If a social robot becomes a stable emotional anchor but may displace human-oriented social practice, then even an ``effective'' anxiolytic design may have unintended developmental and relational costs. This possibility motivates evaluation approaches that treat the robot as part of a family ecology and that explicitly test how social time and social orientation are redistributed when the robot is present versus absent.

From an IDC 2026 \emph{sustainable futures} perspective, we interpret sustainability here as \emph{social sustainability}: supporting children's long-term wellbeing while sustaining caregiving ecologies in everyday home life. For in-home social agents, a sustainable outcome is not only short-term comfort or engagement, but whether benefits transfer to human relationships and whether families can maintain healthy routines without unintended dependency.

\subsection{Conceptual Framework: Engagement Is Not Transfer}
Synthesizing these threads, we propose a conceptual framework that distinguishes \emph{engagement} from \emph{transfer} and clarifies how a robot intended as a scaffold can become a socially central destination in the child's ecology. We organize outcomes into four categories that are often conflated in child-facing social agent evaluations. \emph{Engagement outcomes} (e.g., usability, enjoyment, perceived ease of interaction, attachment, time-on-device) capture whether children can and will interact with the system. \emph{Affective outcomes} (e.g., anxiety reduction, emotional stability) capture whether interaction changes the child's emotional state in beneficial ways. \emph{Transfer outcomes} (e.g., increased human-directed social bids, improved sensitivity to human affect cues, spontaneous real-world empathy) capture whether changes extend beyond the robot-mediated context. \emph{Ecological outcomes} (e.g., displacement/reallocation of social time, caregiver mediation patterns, changes in household routines) capture system-level consequences in the child's relational ecosystem.

We emphasize a distinction that is often blurred in discussion: \emph{substitution-risk signatures} can be inferred from robot-centered orientation and bounded transfer outcomes, whereas \emph{displacement} in the strict sense refers to reallocation of time across humans and devices and typically requires time-use measurement or logging \cite{vandewater2007_time_displacement,rideout2022_media_time_use}. In this study, we prioritize transfer-relevant outcomes and treat displacement as a hypothesis rather than a directly measured quantity. We did not collect time-use diaries or telemetry logs; thus, claims about time reallocation rely on guardian-reported substitution patterns (see Section~\ref{sec:procedure}) and qualitative accounts, which are informative but not definitive measures of displacement.

Within this structure, we define two divergent trajectories. \emph{Social scaffolding} occurs when robot interaction increases subsequent human-directed engagement---for example, when the child initiates more interaction with caregivers or peers and applies emotion understanding in real interpersonal contexts. \emph{Substitution risk} occurs when robot interaction becomes an endpoint: social orientation remains concentrated in the child--robot dyad and does not generalize---or diminishes---in human contexts. We treat this as a risk trajectory that can be signaled by bounded transfer outcomes and caregiver accounts, even without direct evidence of time-use displacement. Critically, both trajectories can coexist with high engagement and high usability; thus, engagement is not a sufficient indicator of scaffolding.

In this paper, we operationalize \emph{transfer} using human-oriented outcomes (SMS, RMET, BES) and qualitative evidence of human-directed reorientation versus siloing (e.g., TRF1--TRF3/WDR3 versus TRF4--TRF5/ENG2 in Supplementary Tables~S19--S20). Under this framing, two contrasting patterns are testable in a withdrawal design. A \emph{scaffolding pattern} would manifest if continued access yields equal-or-better transfer trajectories than withdrawal, and benefits persist and reappear in HHI when the robot is absent. A \emph{substitution-risk pattern} would manifest if continued access yields high engagement and anxiety relief, yet transfer outcomes stagnate or decline relative to withdrawal, with qualitative accounts emphasizing robot-centered routines and limited handoff to caregivers or peers. Displacement in the strict time-use sense is a related but distinct hypothesis that typically requires objective time-allocation measures or usage logging; we treat it as future work (see also Section~\ref{sec:discussion_alt} and Section~\ref{sec:discussion_limitations}).

Finally, we argue that \emph{withdrawal conditions} provide a pragmatic method for testing scaffolding versus substitution risk in real-world contexts. A withdrawal period can reveal how social behaviors reallocate in the robot's absence---though such reallocation may reflect multiple ecological factors, including caregiver co-activity and routine changes, not only the robot's removal itself---suggesting either a scaffolding pattern (benefits persist and reappear in HHI) or a substitution-risk pattern (social orientation concentrated in the child--robot dyad). This framing aligns with recent calls to evaluate effectiveness beyond controlled settings \cite{david2025_lab_to_reality} and with long-term HRI perspectives emphasizing that intervention dynamics unfold over time and within household routines \cite{matheus2025_long_term_interactions_review}. Guided by this framework, our study treats high usability and strong engagement not as endpoints but as context: we examine whether a highly usable consumer-grade social robot can simultaneously produce affective benefits while shaping transfer and ecological outcomes in competing directions. This lens informs both our analysis plan (separating engagement, affective, transfer, and ecological evidence) and our design implications (prioritizing ``handoff'' mechanisms that preserve anxiolytic benefits while promoting human-oriented practice).


\section{Methods}
\label{sec:methods}

\subsection{Study Design Overview}
\label{sec:design}
We conducted an 8-week, two-arm, parallel-group randomized controlled trial (RCT) situated in participants' natural home environments. The study was designed as a \emph{withdrawal-as-transfer-probe} evaluation: after baseline assessment (T1), children were randomized either to (a) \textbf{continued access} to the robot (continued-access group) or (b) \textbf{withdrawal} of the robot from the home (withdrawal group). Outcomes were assessed at baseline (T1; Day 0), midpoint (T2; Day 28), and post-intervention (T3; Day 56). The trial used a mixed-methods approach, combining repeated quantitative measures with semi-structured guardian interviews at T2 and T3 to explain observed trajectories and mechanisms in the family ecology.

Figure~\ref{fig:timeline} provides a schematic of the study flow and timepoints.

\IfFileExists{1.pdf}{%
\begin{figure}[t]
    \centering
    \includegraphics[width=\linewidth]{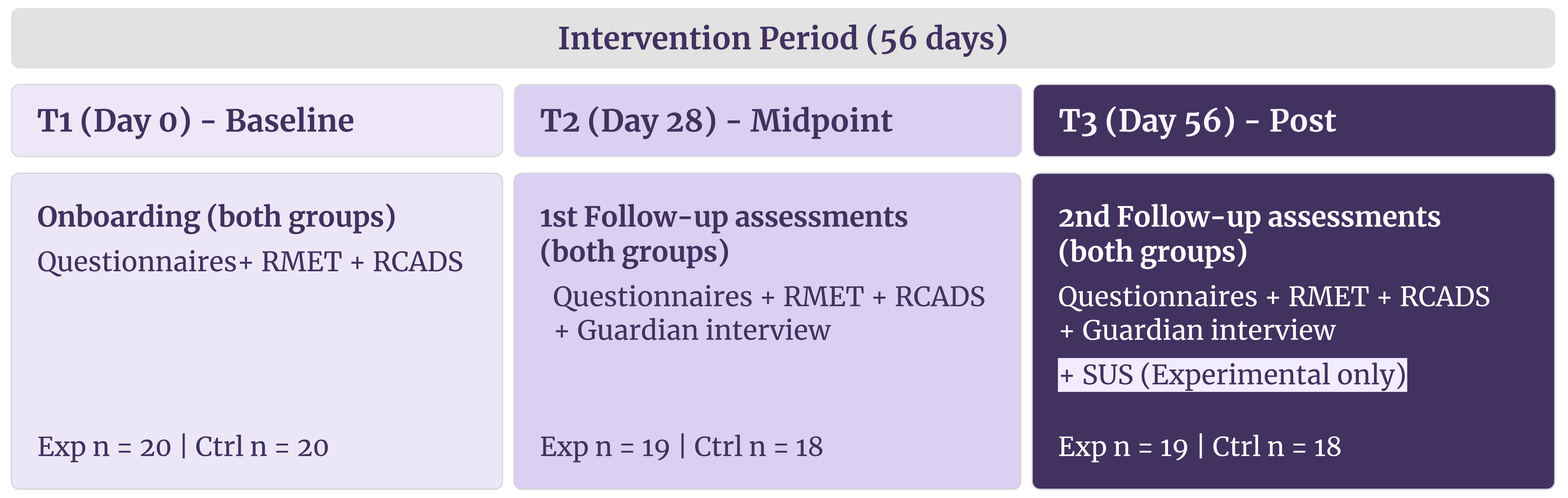}%
    \caption{Study timeline and data collection schedule.}
    \label{fig:timeline}
\end{figure}
}{}

\subsection{Intervention Technology: Qrobot Platform and Home Deployment Context}
\label{sec:qrobot}

\subsubsection{Qrobot as a Commercial Off-the-Shelf Social Robot}
Qrobot is a commercial consumer product originally developed through collaboration between the Shenzhen Institutes of Advanced Technology (Chinese Academy of Sciences), Tencent, and an industry partner. The platform is positioned as a desktop intelligent interactive terminal that supports information services (e.g., news, weather), audio/video playback, and child-oriented educational and entertainment content. The device embodies Tencent's iconic \emph{QQ penguin} character, which provides an immediately legible social form factor for children and families.

\subsubsection{Hardware and Embodiment}
Qrobot is a desktop robot with an integrated display (approximately 1.44 inches), an onboard camera (approximately 1.3 MP), and an internal loudspeaker. It is capable of simple expressive motions (e.g., head rotation) driven by internal motor and gear components. The device also supports direct touch interaction via a capacitive touch sensor (located on the head in the standard configuration) and includes additional sensing elements (e.g., light/gesture-sensitive components) that can trigger behaviors.

\subsubsection{Interaction Modalities and System Architecture}
Qrobot interaction is primarily voice-driven: users issue spoken commands and receive spoken responses, often accompanied by embodied movement and visual output on the robot's display. The platform relies on an application ecosystem that supports multiple functions (e.g., child content, music, quizzes), and can receive updates and extensions via network connectivity.
A characteristic feature of Qrobot is that core services can be coordinated through a companion computer-based management software and cloud-backed services, enabling continuous feature expansion and data synchronization across sessions.

\subsubsection{Study-Relevant Interaction Characteristics}
Because Qrobot is a general-purpose consumer robot rather than a bespoke clinical device, we intentionally evaluated it \emph{as used in everyday family routines}. Families were instructed not to adopt a novel scripted therapy protocol; instead, they were asked to maintain their existing, naturalistic usage patterns in the continued-access group.
Across onboarding and subsequent interviews, the most commonly described child-facing uses relevant to the present outcomes included: (1) interactive voice conversation and prompt-response activities; (2) child-oriented educational games and quizzes; (3) story/audio content; and (4) image- and prompt-based emotion-related learning activities (e.g., labeling feelings in pictures or characters). Because the study imposed no structured protocol and Qrobot does not provide usage logging, specific interaction types, durations, and content were not systematically recorded or objectively verified; the activity categories above are therefore based exclusively on guardian report and should be treated as descriptive characterizations rather than precise usage profiles.

\begin{table*}[t]
\centering
\caption{Qrobot platform characteristics and interaction modalities relevant to the present study.}
\label{tab:qrobot_profile}
\resizebox{0.98\linewidth}{!}{
\begin{tabular}{p{0.22\linewidth} p{0.74\linewidth}}
\hline
\textbf{Dimension} & \textbf{Description (study-relevant)} \\
\hline
Embodiment & Desktop robot with QQ penguin form factor; designed to function as a socially legible companion device in the home. \\
\hline
I/O channels & Voice interaction (primary); audio playback; on-device display output; simple embodied motion (e.g., head movement); touch-triggered interaction. \\
\hline
Sensing & Onboard camera; capacitive touch sensor; additional sensors enabling light/gesture-triggered behaviors (device-dependent configuration). \\
\hline
Content ecosystem & Multi-function application platform including child-oriented content (stories, education, entertainment) and information services; supports updates/extensions via network connectivity and companion management software. \\
\hline
Deployment context & Naturalistic home use; no therapist-led protocol imposed; caregiver may co-use or supervise depending on household routine. \\
\hline
\end{tabular}}
\end{table*}

\subsection{Participants and Recruitment}
\label{sec:participants}
Participants were recruited using a multi-channel strategy combining online outreach (e.g., major social media platforms commonly used in China) and community-based/offline referral. Eligibility screening was conducted with the primary guardian (hereafter ``guardian'').

Children were eligible if they: (1) were between 5 and 10 years of age (enrolled sample ultimately ranged 5--9 years); (2) had a formal clinical diagnosis of Autism Spectrum Disorder (ASD) reported by caregivers; (3) had prior exposure to Qrobot not exceeding three months at enrollment (to ensure familiarity without deep long-term integration); (4) had sufficient receptive language to understand simple instructions; and (5) provided age-appropriate assent with guardian written informed consent. Children were excluded if they had severe sensory or neurological comorbidities that would interfere with safe participation, significant aggressive or self-injurious behaviors that would preclude safe home participation, concurrent participation in other structured social-skills intervention studies in the prior six months, or regular use of other social robot systems. Forty children completed baseline assessment and were randomized (continued access: $n=20$; withdrawal: $n=20$). At T3, analyzable questionnaire/performance data were available for $n=19$ in the continued-access group and $n=18$ in the withdrawal group (see Table~\ref{tab:descriptives}). Participant demographics are provided in Supplementary Table~S3, and baseline characteristics are summarized in Table~\ref{tab:baseline}. To assess robustness to this baseline imbalance, we conducted covariate-adjusted sensitivity analyses that include child age and ASD history months as covariates in the mixed-effects models (Supplementary Tables~S13--S17). Key Group$\times$Time conclusions were unchanged.

\begin{table*}[t]
\centering
\caption{Participant characteristics and baseline balance between groups. SMD = Standardized Mean Difference (absolute values closer to 0 indicate better balance).}
\label{tab:baseline}
\begin{tabular}{lccccc}
\hline
\textbf{Characteristic} & \textbf{Continued access (n=20)} & \textbf{Withdrawal (n=20)} & \textbf{SMD} & \textbf{$t$} & \textbf{$p$} \\
\hline
\multicolumn{6}{l}{\textit{Demographics}} \\
\quad Age (years) & 6.95 (1.15) & 6.60 (1.50) & 0.26 & 0.83 & .412 \\
\quad ASD history (months) & 15.15 (7.94) & 20.80 (8.33) & $-$0.69 & $-$2.20 & .034 \\
\quad Gender (\% male) & 50\% & 55\% & --- & --- & --- \\
\quad Residence (\% urban) & 40\% & 50\% & --- & --- & --- \\
\hline
\multicolumn{6}{l}{\textit{Baseline outcome measures (T1)}} \\
\quad SMS (Social Motivation) & 35.10 (6.50) & 32.80 (6.04) & 0.37 & 1.16 & .254 \\
\quad RMET (Emotion Recognition) & 13.80 (3.27) & 13.75 (3.51) & 0.01 & 0.05 & .963 \\
\quad BES (Empathy) & 37.40 (6.32) & 34.80 (5.98) & 0.42 & 1.34 & .189 \\
\quad SCARED (Anxiety---Guardian) & 37.80 (9.98) & 40.35 (8.80) & $-$0.27 & $-$0.86 & .397 \\
\quad RCADS (Anxiety---Child) & 63.95 (18.38) & 69.20 (15.20) & $-$0.31 & $-$0.98 & .331 \\
\hline
\multicolumn{6}{l}{\footnotesize \textit{Note.} Values are Mean (SD) unless otherwise indicated. We report SMD as a descriptive balance metric.} \\
\multicolumn{6}{l}{\footnotesize ASD history (months) shows a moderate baseline imbalance ($|\mathrm{SMD}| = 0.69$; $p = .034$).}
\end{tabular}
\end{table*}

\begin{figure}[t]
    \centering
    \IfFileExists{2.pdf}{%
        \includegraphics[width=\linewidth]{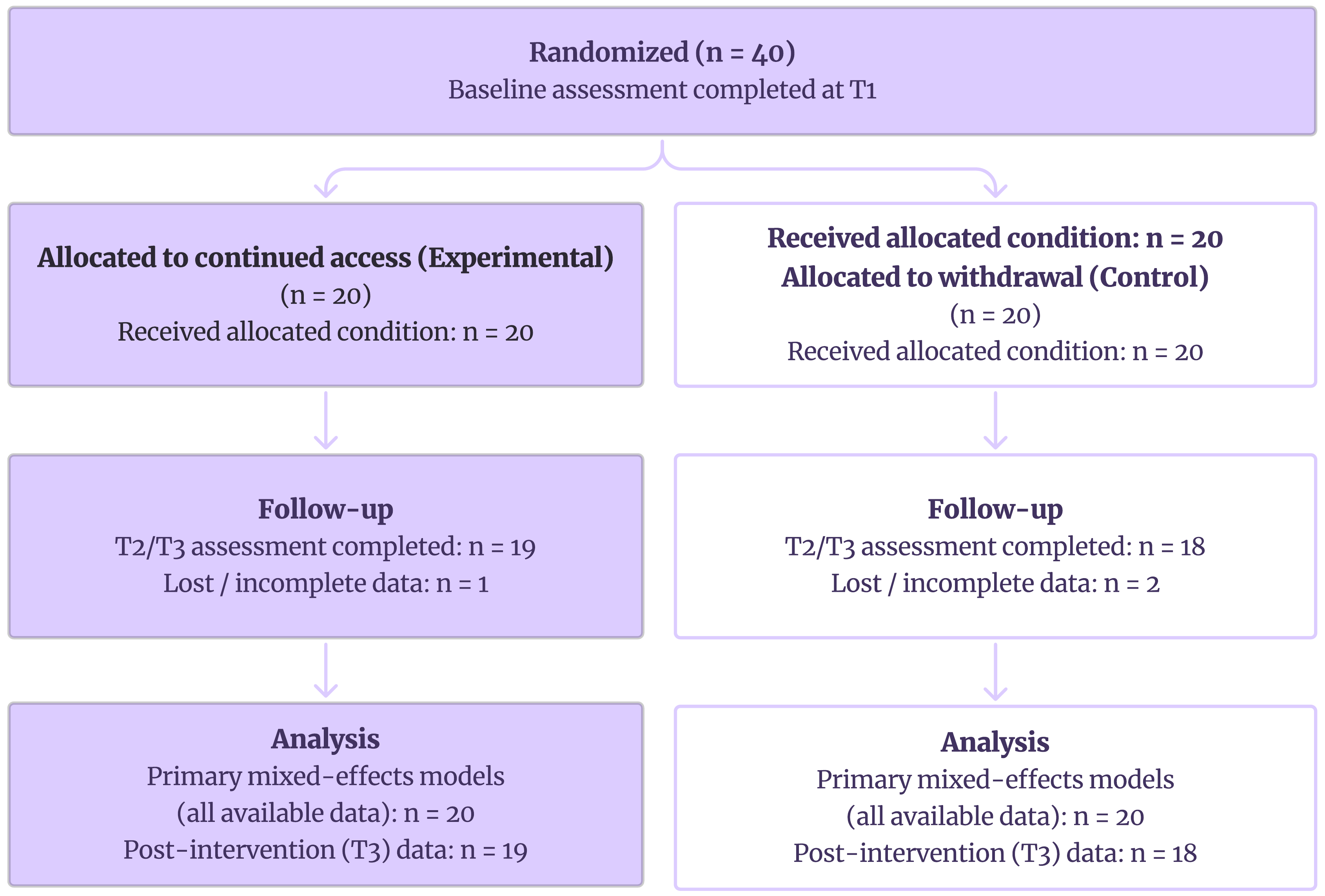}%
    }{}
    \caption{Participant flow diagram (CONSORT-style).}
    \label{fig:consort}
\end{figure}

\subsection{Selection and Participation of Children}
\label{sec:selection_participation_children}

Following IDC norms for reporting research with children, we explicitly describe how children were selected, how participation was supported, and what safeguards were implemented throughout recruitment, withdrawal, and data collection. In this study, children participated as \emph{research participants} (i.e., not as co-designers). Our goal was to minimize burden and risk while preserving ecological validity in home use.

The study protocol was reviewed and approved by the Institutional Review Board at the authors' institution, and all procedures complied with applicable regulations for research involving minors. Families were recruited through community channels and online outreach. Initial screening was conducted with the primary guardian to confirm eligibility, including child age, caregiver-reported clinical ASD diagnosis, prior exposure to Qrobot not exceeding three months, and absence of exclusion criteria that could increase risk or confound outcomes. This selection strategy balanced two constraints: ensuring that children could meaningfully engage with the study activities while avoiding participation demands that could produce undue stress.

Guardians provided written informed consent prior to any data collection, and children were invited to assent using an age-appropriate explanation (e.g., ``We will ask you some questions and play a short picture game. You can stop at any time.''). Assent was treated as an ongoing process rather than a one-time event: at each timepoint, researchers and guardians checked whether the child appeared willing to continue. Children could pause or discontinue participation at any point without penalty, and families could withdraw from the study without providing a reason.

To reduce burden and increase accessibility, study activities were structured to be brief and flexible. When needed, guardians or researchers supported comprehension by reading items aloud using standardized phrasing and neutral tone, avoiding coaching or leading prompts. Breaks were permitted at any time, remote sessions were scheduled around family routines to reduce disruption, and participation occurred in familiar home environments. Because the withdrawal condition required temporary removal of a valued device, we implemented additional safeguards: guardians were briefed in advance that some children might initially ask about the robot or show disappointment, and were provided guidance for supportive responses. Guardians were asked to monitor for distress escalation (e.g., sustained sleep disruption, prolonged dysregulation, or significant functional decline), and families could pause participation and consult the research team if withdrawal appeared to cause sustained distress.

We collected questionnaire responses and audio recordings of guardian interviews, with audio recording occurring only with explicit permission. Interview transcripts were de-identified by removing names and any directly identifying information, and data were stored on access-controlled systems following institutional policy. For families in the withdrawal condition, Qrobot access could be restored after the post-intervention assessment (T3).

\subsection{Randomization and Allocation}
\label{sec:randomization}
Following baseline assessment (T1), participants were randomized in a 1:1 ratio to the continued-access or withdrawal condition. Randomization was generated using a computer-based random sequence and applied only after completion of baseline measures to prevent allocation-driven baseline bias. Allocation was communicated to guardians immediately after baseline completion.

\subsection{Procedure}
\label{sec:procedure}
The study consisted of three phases:

\paragraph{Phase 1: Baseline (T1; Day 0).}
An onboarding session was conducted with each family. Guardians received study instructions, completed consent documentation, and confirmed eligibility. Baseline quantitative measures were then administered to guardians and children (see Section~\ref{sec:measures}). Families in both conditions had Qrobot present at baseline by design (participants had pre-existing access).

\paragraph{Phase 2: Intervention (T1--T3; Day 0--Day 56) with midpoint assessment (T2; Day 28).}
After randomization, families in the continued-access group were instructed to continue their typical Qrobot usage. Families in the withdrawal group were instructed to remove Qrobot from the child's accessible environment (e.g., store the device away, out of sight and out of reach) for the remainder of the study period. Because telemetry was not available for this commercial device (see Section~\ref{sec:discussion_alt}), adherence was assessed via guardian report during scheduled study contacts (T2 and T3), including direct probes about whether Qrobot was accessed during the withdrawal period and where the device was stored. We also asked guardians to describe notable substitutions in daily routines (e.g., increased use of other screen-based media or increased caregiver co-activity), since these ecological shifts can influence transfer outcomes.
At T2 and T3, all 18 analyzable withdrawal-group guardians confirmed that Qrobot remained inaccessible throughout the study period; no unplanned access episodes were reported. Common substitutions described by withdrawal-group guardians included increased tablet/TV screen time ($n = 9$) and more frequent caregiver-led shared activities (e.g., reading, outdoor play; $n = 11$). In the continued-access group, guardians reported daily or near-daily Qrobot use ($n = 16$), with the remainder ($n = 3$) describing use on most days (4--5 times per week). We note that, without device telemetry, usage data for the continued-access group---including interaction duration, frequency, and content type---relies entirely on guardian report and cannot be objectively verified; dosage heterogeneity within this group therefore cannot be ruled out as a source of outcome variability.
At midpoint (T2), guardians completed questionnaires and participated in a semi-structured interview. Children completed age-appropriate tasks and self-report measures with guardian assistance when needed.

\paragraph{Phase 3: Post-intervention (T3; Day 56).}
At the end of the 8-week period, all quantitative measures were re-administered. Guardians also completed a second semi-structured interview reflecting on the full period. Children in the continued-access group completed the System Usability Scale (SUS) to assess perceived usability of the Qrobot interaction.

\begin{table*}[t]
\centering
\caption{Assessment schedule and data sources across timepoints.}
\label{tab:schedule}
\resizebox{0.98\linewidth}{!}{
\begin{tabular}{p{0.22\linewidth} p{0.24\linewidth} p{0.16\linewidth} p{0.16\linewidth} p{0.16\linewidth}}
\hline
\textbf{Construct / Data} & \textbf{Instrument / Source} & \textbf{T1 (Day 0)} & \textbf{T2 (Day 28)} & \textbf{T3 (Day 56)} \\
\hline
Social motivation & Social Motivation Scale (guardian-report) & \checkmark & \checkmark & \checkmark \\
Emotion inference & RMET (Child version) & \checkmark & \checkmark & \checkmark \\
Empathy & Basic Empathy Scale (guardian-report) & \checkmark & \checkmark & \checkmark \\
Anxiety (guardian) & SCARED (guardian-report) & \checkmark & \checkmark & \checkmark \\
Anxiety/depression (child) & RCADS (child self-report) & \checkmark & \checkmark & \checkmark \\
Usability (continued-access only) & SUS (child-rated) & -- & -- & \checkmark \\
Qualitative experience & Semi-structured interview (guardian) & -- & \checkmark & \checkmark \\
\hline
\end{tabular}}
\end{table*}

\subsection{Quantitative Measures}
\label{sec:measures}
We selected measures to distinguish (1) engagement/usability, (2) affect regulation, and (3) human-oriented transfer-relevant outcomes.
Given the young age range (5--9), we used guardian-report instruments (SMS, BES, SCARED) as primary indicators for their respective constructs and administered child-facing measures (RMET; RCADS; SUS) with standardized, neutral support when needed. We treat RCADS as a secondary, exploratory child self-report indicator and interpret it cautiously for younger participants. All instruments were administered in Mandarin Chinese using validated translations. Internal consistency (Cronbach's $\alpha$) was computed at baseline (T1): SMS $\alpha = .84$; BES $\alpha = .81$; SCARED $\alpha = .88$; RCADS $\alpha = .86$. These values are consistent with published psychometrics for the respective Chinese versions \cite{wang2013_scared_chinese,guan2012_rcads_chinese}. RMET does not yield an internal consistency coefficient but was scored following the standardized protocol for the child version \cite{baroncohen2001_rmet_child}.

\textit{Guardian-report measures.} Guardians completed the Social Motivation Scale (SMS), a 20-item instrument assessing the child's drive to seek, maintain, and enjoy social interaction, with items rated on a Likert scale and summed to produce a total score (higher scores indicating greater social motivation). Guardians also completed the Basic Empathy Scale (BES), capturing cognitive and affective empathy dimensions, and the Screen for Child Anxiety Related Emotional Disorders (SCARED), a widely used screening instrument for anxiety symptoms and subtypes.

\textit{Child performance and self-report measures.} Children completed the Reading the Mind in the Eyes Test (RMET), Child Version, which assesses the ability to infer mental/emotional states from images focused on the eye region; because validated parallel forms are not available, repeated administrations may be subject to retest effects (see Section~\ref{sec:discussion_limitations}); we therefore prioritize between-group divergence over within-group absolute change when interpreting RMET trajectories. Children also completed the Revised Children's Anxiety and Depression Scale (RCADS; Chinese version \cite{guan2012_rcads_chinese}) as a secondary, exploratory self-report indicator. For younger children and/or those requiring support, items were read aloud by a trained researcher using a standardized script: each item was read verbatim in a neutral tone, without rephrasing or explaining content; if the child asked for clarification, the researcher repeated the item once and encouraged the child to respond based on their own understanding. Across both groups at T1, 28 of 40 children (70\%) received read-aloud support; this was more common in the 5--6 age band (17/17, 100\%) than in the 7--9 age band (11/23, 48\%). Guardian presence was permitted during RCADS administration for child comfort, but guardians were instructed not to interpret or suggest responses.

\textit{Usability assessment.} At T3, children in the continued-access group completed the System Usability Scale (SUS) to evaluate perceived ease of use of the Qrobot system. Given the young age range, SUS items were administered with standardized, neutral read-aloud support when needed. SUS scoring followed the standard 0--100 conversion procedure, and reverse-worded items were reverse-coded so that higher values consistently indicate better usability.

\begin{table*}[t]
\centering
\caption{Quantitative measures: respondent, format, and scoring.}
\label{tab:measures}
\resizebox{0.98\linewidth}{!}{
\begin{tabular}{p{0.14\linewidth} p{0.18\linewidth} p{0.18\linewidth} p{0.18\linewidth} p{0.26\linewidth}}
\hline
\textbf{Construct} & \textbf{Measure} & \textbf{Respondent} & \textbf{Scale / Output} & \textbf{Notes (administration/scoring)} \\
\hline
Social motivation & SMS & Guardian & Total + subscales & Likert; summed scores (higher = greater social motivation) \\
Emotion inference & RMET (Child) & Child & Correct responses (0--28) & Performance task; standardized instructions \\
Empathy & BES & Guardian & Total + subscales & Cognitive + affective empathy \\
Anxiety & SCARED & Guardian & Total + subscales & Screening; compare change over time \\
Anxiety/depression & RCADS & Child & Total + subscales & Self-report (secondary/exploratory); read-aloud support when needed \\
Usability & SUS & Child (Exp only) & 0--100 & Standard SUS conversion algorithm \\
\hline
\end{tabular}}
\end{table*}

\subsection{Qualitative Data Collection: Semi-Structured Interviews}
\label{sec:qual}
Guardians participated in semi-structured interviews at T2 and T3 to capture lived experiences and mechanism-relevant observations. The interview protocol moved from global impressions (integration into daily life, acceptability) to targeted probes aligned with study outcomes (social initiation/maintenance/enjoyment; emotion recognition and empathic responding; emotional stability and anxiety). The protocol emphasized concrete behavioral examples (e.g., ``Describe a specific time your child initiated interaction with another person'') to reduce abstraction and retrospective bias.

Interviews were conducted remotely via Tencent Meeting and were audio-recorded with permission. Recordings were transcribed verbatim within 24 hours and de-identified.
Interviews also included probes relevant to the withdrawal-as-transfer-probe design, such as how the robot was integrated into routines (or removed during withdrawal), whether children requested the robot during the withdrawal period, and whether caregivers observed changes in human-directed social bids, co-use patterns, or substitutions with other media/devices.
A total of 74 interviews were completed across both timepoints (T2: $n = 37$; T3: $n = 37$), with interview durations ranging from 18 to 42 minutes (median = 26 minutes). All interviews were conducted in Mandarin Chinese. Quotes presented in this paper were translated into English by bilingual members of the research team, with a second bilingual researcher reviewing each translation for semantic fidelity and tone; discrepancies were resolved through discussion.

\subsection{Analysis Plan}
\label{sec:analysis}

\subsubsection{Quantitative Analysis}
We first computed descriptive statistics (M, SD) for each measure by group and timepoint. To test longitudinal group differences while accounting for within-participant correlation, we modeled each outcome using linear mixed-effects models with fixed effects for \texttt{Group}, \texttt{Time} (T1/T2/T3), and their interaction (\texttt{Group}$\times$\texttt{Time}), and a random intercept for participant. We additionally report post-intervention (T3) between-group comparisons using independent samples $t$-tests with Holm correction for multiple outcomes (Table~\ref{tab:main_results}). Effect sizes (Cohen's $d$) and 95\% confidence intervals are reported alongside $p$-values.
As a robustness check motivated by baseline imbalance in ASD history (months), we additionally conducted covariate-adjusted sensitivity analyses that include child age and ASD history months as covariates in the mixed-effects models (Supplementary Tables~S13--S17).

\subsubsection{Qualitative Analysis}
Interview transcripts were analyzed using thematic analysis with a hybrid deductive--inductive coding strategy. A priori domains were derived from the study framework (engagement/attachment; anxiety/regulation; transfer to HHI; emotion generalization; withdrawal ecology). Within these domains, codes were refined iteratively through repeated reading and constant comparison.

Coding was conducted at the meaning-unit level (typically 1--3 sentences per unit), allowing multi-coding when a segment reflected multiple mechanisms (e.g., ``robot as refuge'' and ``low human transfer''). An initial subset of transcripts was open-coded to refine operational definitions and inclusion/exclusion criteria for each code. We maintained an audit trail documenting codebook revisions and analytic memos, and disagreements were resolved through discussion and consensus. Two coders (a doctoral researcher with training in qualitative methods and child development, and a research assistant with HRI background) independently coded all transcripts. To quantify reliability, we computed Cohen's $\kappa$ on a randomly selected subset of 20 transcripts (27\% of total); inter-rater agreement was substantial ($\kappa = .78$). Discrepancies were then resolved through discussion (and, when necessary, consultation with the senior author) to produce a consensus-coded dataset used for analysis.

To avoid ``story-first'' interpretation, we explicitly searched for counterexamples and mixed cases (e.g., families describing anxiety relief without transfer, transfer gains without anxiety change, or substitutions to other devices during withdrawal) and retained these segments as negative cases. Negative cases were documented in analytic memos linked to individual transcripts and reviewed during weekly coding meetings to ensure they informed theme refinement rather than being dismissed. Translation followed a bilingual-review procedure: quotes were translated by one bilingual researcher and reviewed by a second; discrepancies were adjudicated through discussion until consensus was reached.

\subsubsection{Mixed-Methods Integration}
We integrated quantitative and qualitative findings through joint displays aligning: (1) engagement/usability outcomes, (2) affective/anxiety outcomes, and (3) transfer-relevant human-oriented outcomes. Qualitative themes were used to explain the mechanisms behind diverging trajectories, with particular attention to (a) evidence of ``handoff'' from robot-mediated skills to HHI, and (b) evidence of ``siloing'' where interaction remained confined to the child--robot dyad.


\section{Results}
\label{sec:results}
We report descriptive statistics by condition and timepoint in Table~\ref{tab:descriptives}. Our primary longitudinal inference is based on linear mixed-effects models with fixed effects for \texttt{Group}, \texttt{Time}, and their interaction (\texttt{Group}$\times$\texttt{Time}) and a random intercept for participant (see Methods; full parameters in Supplementary Tables~S6--S10). For interpretability, we also report follow-up post-intervention (T3) between-group comparisons with effect sizes in Table~\ref{tab:main_results}.
To orient these results to our research questions: Sections~\ref{sec:results_engagement}--\ref{sec:results_transfer} address \textbf{RQ1} by separating engagement/usability, anxiety, and transfer outcomes; Sections~\ref{sec:results_transfer} and \ref{sec:results_integration} address \textbf{RQ2} by characterizing post-withdrawal reallocation (``handoff'' versus ``siloing''); and \textbf{RQ3} is developed in the Discussion through evaluation and design implications (Sections~\ref{sec:discussion_evaluation} and \ref{sec:discussion_design}).
Baseline assessment included $N=40$ children ($n=20$ per condition). At T3, analyzable questionnaire/performance data were available for $n=19$ in the continued-access (robot access) group and $n=18$ in the withdrawal group (Table~\ref{tab:descriptives}). Full participant demographics are provided in Supplementary Table~S3.

\begin{table*}[t]
\centering
\caption{Descriptive statistics for primary outcomes by group and timepoint. Higher SMS, RMET, and BES scores indicate better functioning; lower SCARED and RCADS scores indicate less anxiety.}
\label{tab:descriptives}
\resizebox{\linewidth}{!}{
\begin{tabular}{llcccccc}
\hline
& & \multicolumn{3}{c}{\textbf{Continued-access group}} & \multicolumn{3}{c}{\textbf{Withdrawal group}} \\
\cmidrule(lr){3-5} \cmidrule(lr){6-8}
\textbf{Outcome} & \textbf{Direction} & \textbf{T1} & \textbf{T2} & \textbf{T3} & \textbf{T1} & \textbf{T2} & \textbf{T3} \\
& & \textbf{(Day 0)} & \textbf{(Day 28)} & \textbf{(Day 56)} & \textbf{(Day 0)} & \textbf{(Day 28)} & \textbf{(Day 56)} \\
\hline
\multicolumn{8}{l}{\textit{Transfer outcomes (human-oriented)}} \\
\quad SMS & $\uparrow$ better & 35.10 (6.50) & 32.32 (6.19) & 30.21 (5.92) & 32.80 (6.04) & 35.67 (6.27) & 41.17 (6.60) \\
\quad RMET & $\uparrow$ better & 13.80 (3.27) & 13.63 (3.29) & 13.26 (3.54) & 13.75 (3.51) & 14.89 (2.78) & 16.61 (2.57) \\
\quad BES & $\uparrow$ better & 37.40 (6.32) & 36.16 (5.96) & 34.89 (5.69) & 34.80 (5.98) & 37.33 (6.67) & 40.61 (7.01) \\
\hline
\multicolumn{8}{l}{\textit{Affective outcomes (anxiety)}} \\
\quad SCARED & $\downarrow$ better & 37.80 (9.98) & 35.37 (10.25) & 32.58 (9.38) & 40.35 (8.80) & 39.06 (7.52) & 40.17 (6.92) \\
\quad RCADS & $\downarrow$ better & 63.95 (18.38) & 45.58 (13.50) & 40.21 (12.89) & 69.20 (15.20) & 67.72 (12.53) & 68.83 (11.91) \\
\hline
& & $n=20$ & $n=19$ & $n=19$ & $n=20$ & $n=18$ & $n=18$ \\
\hline
\multicolumn{8}{l}{\footnotesize \textit{Note.} Values are Mean (SD). SMS = Social Motivation Scale; RMET = Reading the Mind in the Eyes Test;} \\
\multicolumn{8}{l}{\footnotesize BES = Basic Empathy Scale; SCARED = Screen for Child Anxiety Related Emotional Disorders;} \\
\multicolumn{8}{l}{\footnotesize RCADS = Revised Children's Anxiety and Depression Scale.}
\end{tabular}}
\end{table*}

\subsection{Engagement and usability outcomes (high-usability context)}
\label{sec:results_engagement}
\begin{figure*}[t]
    \centering
    \includegraphics[width=\linewidth]{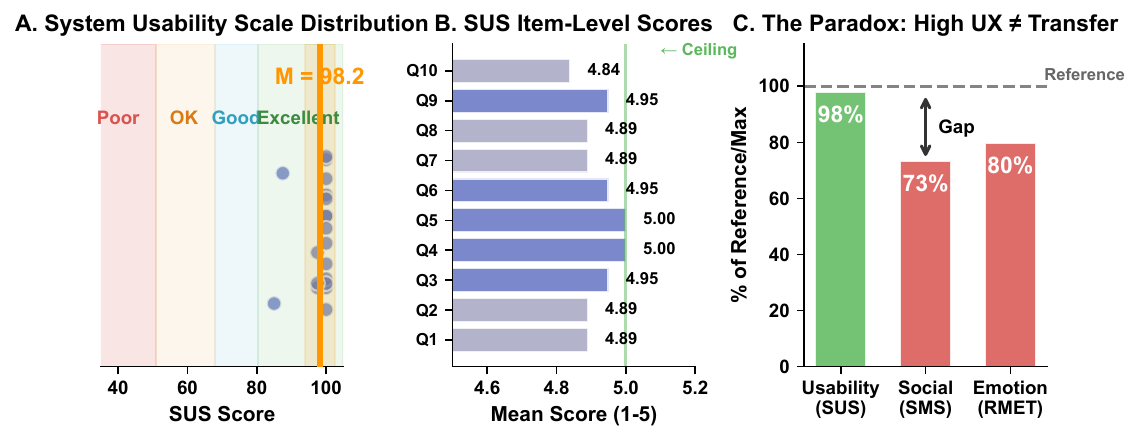}
    \caption{Usability and engagement context for interpreting downstream outcomes. (A) Distribution of SUS total scores (T3; continued-access group). (B) SUS item-level mean scores (1--5). Panels (C--D) contextualize the ``high usability $\neq$ transfer'' paradox and the scaffold-versus-substitute pathways.}
    \label{fig:usability_paradox}
\end{figure*}

Children in the continued-access condition reported very high usability at post-intervention (T3). The mean System Usability Scale (SUS) score was 98.16 (SD = 4.32; $n=19$), with most scores clustered near the ceiling and item-level means close to 5 (Figure~\ref{fig:usability_paradox}, panels A--B). This indicates that the interaction experience itself was highly accessible and unlikely to be limited by interface friction.

Qualitative interviews converged on the same premise: guardians frequently described Qrobot as a highly preferred, routine-integrated companion in the home. For example, one mother noted:
\begin{quote}
\textit{My child's relationship with Qrobot is very close; he spends a lot of time with it every day.} (P4)
\end{quote}

\subsection{Affective outcomes: anxiety reduction and emotional stability}
\label{sec:results_anxiety}
Consistent with the robot's role as an affective regulator, anxiety outcomes showed clear benefits for continued access. On guardian-reported SCARED, the continued-access group decreased over time (T1: 37.80; T3: 32.58), whereas the withdrawal group remained approximately stable (T1: 40.35; T3: 40.17; Table~\ref{tab:descriptives}). As a follow-up post-intervention contrast at T3, SCARED scores were significantly lower in the continued-access group than in the withdrawal group ($t(35) = -2.79$, $p_{\mathrm{adj}} = .017$, $d = -0.92$; Table~\ref{tab:main_results}), indicating reduced anxiety symptoms.
In the longitudinal mixed-effects model (Supplementary Table~S9; $N_{\mathrm{obs}} = 114$), the Group$\times$Time interactions were significant at T2 ($\beta = -2.64$, $p = .008$, 95\% CI [$-4.60$, $-0.69$]) and T3 ($\beta = -6.54$, $p < .001$, 95\% CI [$-8.50$, $-4.59$]), indicating greater reductions under continued access relative to withdrawal.

\begin{figure*}[t]
    \centering
    \includegraphics[width=\linewidth]{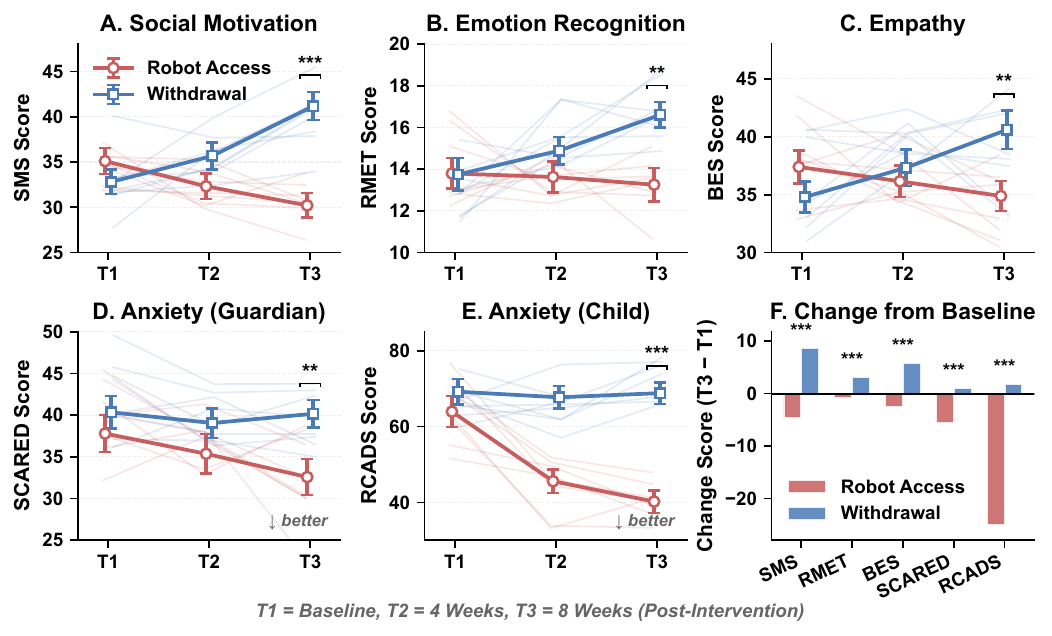}
    \caption{Enhanced trajectories across outcome domains for continued access (robot access) versus withdrawal. Panels D--E show anxiety outcomes (SCARED and RCADS; lower is better).}
    \label{fig:outcome_trajectories_enhanced}
\end{figure*}

As a secondary, exploratory child self-report indicator, RCADS showed a large divergence. In the longitudinal mixed-effects model (Supplementary Table~S10; $N_{\mathrm{obs}} = 114$), the Group$\times$Time interactions were significant at T2 ($\beta = -18.70$, $p < .001$, 95\% CI [$-27.70$, $-9.69$]) and T3 ($\beta = -25.18$, $p < .001$, 95\% CI [$-34.18$, $-16.17$]), indicating greater reductions under continued access relative to withdrawal. As a follow-up post-intervention contrast at T3, RCADS was substantially lower under continued access ($t(35) = -7.00$, $p_{\mathrm{adj}} < .001$, $d = -2.30$; Table~\ref{tab:main_results}). Given the young age range, we interpret RCADS cautiously and foreground guardian-report SCARED for primary anxiety inference. Trajectories are visualized in Figure~\ref{fig:outcome_trajectories_enhanced} (panels D--E).
Child self-report feasibility and missingness were age-sensitive. At T2 and T3, analyzable RCADS data were available for 19/20 (95\%) in the continued-access group and 18/20 (90\%) in the withdrawal group; missingness was concentrated in the continued-access 5--6-year band (5/6 available) and the withdrawal 7--9-year band (9/11 available). Age-restricted sensitivity analyses suggested the pattern persisted when restricting to children aged $\geq 7$ (Group$\times$T3 $\beta = -21.45$, $p = .001$, 95\% CI [$-34.46$, $-8.43$]; Supplementary Table~S18). When restricting to $\geq 8$ (small subsample), estimates remained directionally consistent but were imprecise (Group$\times$T3 $\beta = -22.81$, $p = .096$; Supplementary Table~S18), underscoring limited power for older-only subsamples.

Qualitative reports aligned with these trends. Guardians in the continued-access group described modest but meaningful improvements in emotional regulation (e.g., shorter tantrums) and post-interaction mood carryover:
\begin{quote}
\textit{Before, if something upset him, he might cry and scream for over half an hour, but now he can calm down in about twenty minutes.} (P6)
\end{quote}
\begin{quote}
\textit{After completing the robot-guided activities, his emotional state would be quite good, and this positive mood could last for a while.} (P11)
\end{quote}

Importantly, withdrawal did not appear to trigger sustained distress in qualitative accounts; guardians often described an initial adjustment followed by stability:
\begin{quote}
\textit{At the beginning, the child did ask a few times where the robot went, but he adapted after about a week... overall, there hasn't been much change.} (P29)
\end{quote}

\begin{table*}[t]
\centering
\caption{Follow-up post-intervention (T3) between-group comparisons with effect sizes and 95\% confidence intervals. These $t$-tests provide simple-effect contrasts complementing the primary longitudinal mixed-effects models (Group$\times$Time). All comparisons use independent samples $t$-tests with Holm correction for multiple outcomes.}
\label{tab:main_results}
\resizebox{\linewidth}{!}{
\begin{tabular}{lcccccccc}
\hline
& \multicolumn{2}{c}{\textbf{Continued access}} & \multicolumn{2}{c}{\textbf{Withdrawal}} & & & & \\
\cmidrule(lr){2-3} \cmidrule(lr){4-5}
\textbf{Outcome} & $n$ & M (SD) & $n$ & M (SD) & \textbf{$t$(35)} & \textbf{$p$} & \textbf{$p_{\mathrm{adj}}$} & \textbf{Cohen's $d$ [95\% CI]} \\
\hline
\multicolumn{9}{l}{\textit{Transfer outcomes}} \\
\quad SMS & 19 & 30.21 (5.92) & 18 & 41.17 (6.60) & $-$5.32 & $<$.001 & $<$.001*** & $-$1.75 [$-$2.51, $-$0.99] \\
\quad RMET & 19 & 13.26 (3.54) & 18 & 16.61 (2.57) & $-$3.28 & .002 & .007** & $-$1.08 [$-$1.77, $-$0.39] \\
\quad BES & 19 & 34.89 (5.69) & 18 & 40.61 (7.01) & $-$2.73 & .010 & .017* & $-$0.90 [$-$1.57, $-$0.22] \\
\hline
\multicolumn{9}{l}{\textit{Affective outcomes}} \\
\quad SCARED & 19 & 32.58 (9.38) & 18 & 40.17 (6.92) & $-$2.79 & .009 & .017* & $-$0.92 [$-$1.59, $-$0.24] \\
\quad RCADS & 19 & 40.21 (12.89) & 18 & 68.83 (11.91) & $-$7.00 & $<$.001 & $<$.001*** & $-$2.30 [$-$3.13, $-$1.47] \\
\hline
\multicolumn{9}{l}{\textit{Engagement outcome (continued-access only)}} \\
\quad SUS & 19 & 98.16 (4.32) & --- & --- & --- & --- & --- & --- \\
\hline
\multicolumn{9}{l}{\footnotesize \textit{Note.} $p_{\mathrm{adj}}$ = Holm-corrected $p$-value. Negative $d$ indicates Continued access $<$ Withdrawal.} \\
\multicolumn{9}{l}{\footnotesize For transfer outcomes, negative $d$ = lower (worse) scores under continued access (substitution-risk signature; not a direct displacement measure).} \\
\multicolumn{9}{l}{\footnotesize For affective outcomes, negative $d$ = lower (better) scores under continued access (anxiety reduction).} \\
\multicolumn{9}{l}{\footnotesize *$p < .05$, **$p < .01$, ***$p < .001$ (Holm-corrected).}
\end{tabular}}
\end{table*}

\subsection{Transfer outcomes: human-directed social motivation and emotion understanding}
\label{sec:results_transfer}
In contrast to the anxiolytic benefit of continued access, transfer-relevant outcomes moved in the opposite direction. As shown in Table~\ref{tab:descriptives} and Figure~\ref{fig:outcome_trajectories_enhanced} (panels A--C), the withdrawal group exhibited increases over time in social motivation, emotion inference, and empathy-related behavior, whereas the continued-access group showed flat-to-declining trajectories on these human-oriented measures.

Two of these transfer outcomes (SMS and BES) are guardian-reported, and the primary anxiety outcome (SCARED) is also guardian-reported. While guardian report improves feasibility for young children, it can amplify expectancy and shared-method variance; we therefore interpret transfer trajectories in conjunction with the child performance measure (RMET) and with concrete behavioral examples from guardian interviews.

For \textit{social motivation} (SMS), scores decreased under continued access (T1: 35.10; T3: 30.21) but increased under withdrawal (T1: 32.80; T3: 41.17; Table~\ref{tab:descriptives}). In the longitudinal mixed-effects model (Supplementary Table~S6; $N_{\mathrm{obs}} = 114$), the Group$\times$Time interactions were significant at T2 ($\beta = -5.82$, $p < .001$, 95\% CI [$-8.04$, $-3.61$]) and T3 ($\beta = -13.43$, $p < .001$, 95\% CI [$-15.64$, $-11.22$]), indicating a divergent trajectory favoring withdrawal over time. As a follow-up post-intervention contrast at T3, SMS was significantly lower in the continued-access group than in the withdrawal group ($t(35) = -5.32$, $p_{\mathrm{adj}} < .001$, $d = -1.75$; Table~\ref{tab:main_results}).

For \textit{emotion inference} (RMET), scores improved under withdrawal (T1: 13.75; T3: 16.61) while remaining flat-to-slightly lower under continued access (T1: 13.80; T3: 13.26; Table~\ref{tab:descriptives}). At T3, RMET was significantly lower in the continued-access group ($t(35) = -3.28$, $p_{\mathrm{adj}} = .007$, $d = -1.08$; Table~\ref{tab:main_results}). In the longitudinal mixed-effects model (Supplementary Table~S7; $N_{\mathrm{obs}} = 114$), the Group$\times$Time interactions were significant at T2 ($\beta = -1.82$, $p < .001$, 95\% CI [$-2.65$, $-1.00$]) and T3 ($\beta = -3.91$, $p < .001$, 95\% CI [$-4.74$, $-3.09$]), indicating increasing divergence favoring withdrawal over time.

For \textit{empathy-related behavior} (BES), scores decreased under continued access (T1: 37.40; T3: 34.89) but increased under withdrawal (T1: 34.80; T3: 40.61; Table~\ref{tab:descriptives}). At T3, BES was significantly lower in the continued-access group ($t(35) = -2.73$, $p_{\mathrm{adj}} = .017$, $d = -0.90$; Table~\ref{tab:main_results}). In the longitudinal mixed-effects model (Supplementary Table~S8; $N_{\mathrm{obs}} = 114$), the Group$\times$Time interactions were significant at T2 ($\beta = -3.92$, $p < .001$, 95\% CI [$-5.13$, $-2.71$]) and T3 ($\beta = -8.46$, $p < .001$, 95\% CI [$-9.67$, $-7.25$]), indicating a divergent trajectory favoring withdrawal over time.

Qualitative interviews provided converging mechanism-relevant evidence for these diverging transfer trajectories. In the withdrawal group, guardians commonly described a reallocation of social bids toward caregivers and peers after an initial adjustment to the robot's absence:
\begin{quote}
\textit{From the third week on, he started to seek more interaction with us... he would proactively bring a picture book, sit next to us, and ask, `Mommy, read.' ... he now observes other children at the playground and occasionally tries to join them.} (P25)
\end{quote}
In contrast, guardians in the continued-access group often described robot-centered engagement that did not readily generalize to human contexts:
\begin{quote}
\textit{When guests visit, he still retreats to a corner to play with the robot and doesn't proactively greet them or show them what he learned from the robot.} (P4)
\end{quote}

For emotion understanding, withdrawal-group accounts included spontaneous real-world empathic responding:
\begin{quote}
\textit{She ... brought back her little blanket, covered me with it, and said, `Mommy, rest.'} (P28)
\end{quote}
Whereas continued-access accounts more often described brittle, context-bound learning that remained confined to robot prompts:
\begin{quote}
\textit{He can accurately answer the robot's questions about ... emotions, but in real life, when faced with his younger sister's crying, he often seems confused or indifferent.} (P12)
\end{quote}
These themes are reflected in the qualitative coding framework and detailed codebook (Supplementary Table~S19 and Table~S20).

\subsection{Mixed-methods integration: joint display of outcomes and mechanisms}
\label{sec:results_integration}
Table~\ref{tab:joint_display} integrates quantitative outcomes across domains with qualitative themes that help explain why high engagement and anxiety reduction co-occurred with weaker human-directed transfer under continued access.

\begin{table*}[t]
\centering
\caption{Joint display aligning quantitative outcomes with qualitative themes. Code labels refer to Supplementary Table~S19.}
\label{tab:joint_display}
\resizebox{\linewidth}{!}{
\begin{tabular}{p{0.17\linewidth} p{0.32\linewidth} p{0.33\linewidth} p{0.14\linewidth}}
\toprule
\textbf{Outcome domain} & \textbf{Quantitative signal} & \textbf{Qualitative themes (examples)} & \textbf{Interpretation} \\
\midrule
Engagement / usability & SUS very high at T3 (M = 98.16; Fig.~\ref{fig:usability_paradox}) & ENG1/ENG3: robot as preferred companion; high centrality in routines (e.g., P4) & Not a UX failure; engagement $\neq$ transfer \\
Anxiety / regulation & SCARED and RCADS lower under continued access at T3 (Table~\ref{tab:main_results}; Fig.~\ref{fig:outcome_trajectories_enhanced}, D--E) & ANX1/ANX2: shorter tantrums; mood carryover (P6, P11); ANX3: withdrawal stability (P29) & Affective benefit under continued access \\
Transfer (human-oriented) & SMS, RMET, BES lower under continued access at T3 (Table~\ref{tab:main_results}; Fig.~\ref{fig:outcome_trajectories_enhanced}, A--C) & TRF1--3/WDR3: post-withdrawal human reorientation (P25); TRF4--5: siloing/failed mediation (P4, P9); EMO1 vs EMO3: real-world empathy vs rote labeling (P28, P12) & Substitution risk; withdrawal reveals handoff \\
\bottomrule
\end{tabular}}
\end{table*}

\subsection{Exploratory trade-off pattern: anxiety relief versus transfer outcomes}
\label{sec:results_tradeoff}
\begin{figure*}[t]
    \centering
    \includegraphics[width=\linewidth]{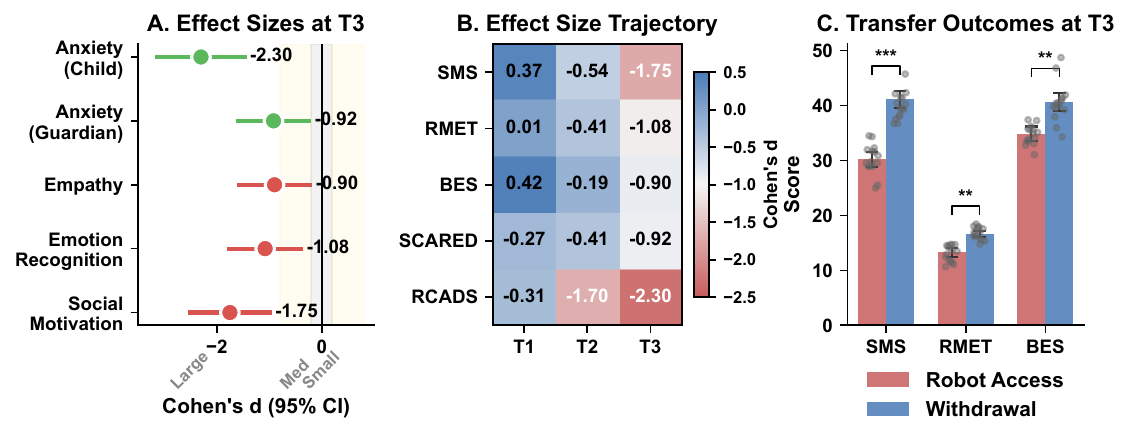}
    \caption{Effect size analysis illustrating the cross-domain trade-off pattern. (A) Cohen's $d$ at T3 with 95\% confidence intervals. (B) Effect-size trajectories across time. (C) Transfer outcomes at T3. (D) Statistical summary. For anxiety outcomes, negative $d$ indicates lower (better) scores under continued access; for transfer outcomes, negative $d$ indicates lower (worse) human-oriented scores under continued access.}
    \label{fig:effect_sizes_tradeoff}
\end{figure*}

To summarize cross-domain patterns in a comparable scale, we visualized standardized effect sizes across time (Figure~\ref{fig:effect_sizes_tradeoff}). Effect sizes were near zero at baseline (T1) and diverged by T3, producing a coherent trade-off: continued access strongly benefited anxiety outcomes (lower SCARED/RCADS; Table~\ref{tab:main_results}) while coinciding with poorer human-oriented transfer outcomes (lower SMS/RMET/BES; Table~\ref{tab:main_results}). We treat this visualization as descriptive and interpret it in conjunction with the withdrawal-as-transfer-probe design and qualitative mechanisms (Table~\ref{tab:joint_display}).

\section{Discussion}
\label{sec:discussion}

This study set out to test a common assumption in child-facing HRI for autism: that sustained, engaging robot interaction functions as a \emph{social scaffold} that supports later human--human interaction (HHI). Instead, our mixed-method evidence foregrounds a more complicated and design-relevant reality. In a naturalistic home deployment of a \emph{highly usable} consumer robot (SUS $\approx$ 98), continued access produced robust \emph{affective benefits} (lower SCARED; RCADS showed a similar direction as a secondary/exploratory child self-report), while \emph{human-oriented transfer outcomes} moved in the opposite direction: social motivation and measures related to emotion understanding/empathy improved more under withdrawal than under continued access. These results sharpen an evaluation blind spot raised in recent syntheses: engagement and feasibility are often treated as success criteria, yet they do not necessarily imply generalization to everyday social life \cite{kouroupa2022_socialrobots_meta,salimi2021_rcts_review,alabdulkareem2022_robot_assisted_therapy_review,david2025_lab_to_reality}. Below, we interpret the findings through our conceptual lens (\emph{Engagement is Not Transfer}), discuss plausible mechanisms that connect comfort to substitution risk, and extract design and evaluation implications for child-facing social agents deployed in homes.

\subsection{Engagement is Not Transfer: a high-usability failure mode}
\label{sec:discussion_paradox}
A central contribution of this work is empirical evidence that \emph{excellent interaction experience} can coexist with \emph{weak transfer}. Children in the continued-access group rated Qrobot as very easy to use (near-ceiling SUS), and caregivers described strong preference and routine integration (ENG1--ENG3). If negative outcomes were driven by interface friction, breakdowns, or usability barriers, we would expect the opposite pattern---i.e., lower usability alongside poorer outcomes. Instead, the high-usability context strengthens a more consequential claim: for child-facing social agents, optimizing engagement and interaction smoothness may be \emph{insufficient} and, under some conditions, may even be \emph{orthogonal} to transfer-relevant goals. This reframing aligns with broader calls to move ``from lab to reality'' and to evaluate how interventions reshape families' lived routines and interaction ecologies over time, not only whether a child can perform well during robot-present sessions \cite{david2025_lab_to_reality,matheus2025_long_term_interactions_review}.

Our results suggest that the scaffolding assumption should be treated as a \emph{testable hypothesis} rather than a default narrative. Recent work has emphasized the potential of robots to support higher-order social cognition and to serve as catalysts for HHI \cite{ghiglino2023artificial,ghiglino2025tom,chen2025catalysts}. However, evidence for durable generalization remains heterogeneous, and follow-up research highlights that post-intervention trajectories depend on how the system is embedded into daily life and mediated by caregivers \cite{ghiglino2021followup,piccolo2024_parental_involvement_review,amirova2023_parental_involvement_ra_at}. Our study adds a complementary and cautionary datapoint: without interaction structures explicitly designed for handoff, a highly engaging consumer robot may settle into a stable niche in the home that does not automatically promote human-directed growth.

\subsection{Withdrawal as a transfer probe: reallocation reveals ``handoff'' versus ``destination''}
\label{sec:discussion_withdrawal}
Methodologically, our withdrawal condition functions as more than a conventional control. It operates as a \emph{transfer probe} that can make visible what robot-present evaluations often obscure: how children's social bids and emotion-attunement behaviors \emph{reallocate} when the robot is not available, though such reallocation may reflect multiple ecological factors (see Section~\ref{sec:discussion_alt} for alternative explanations). The withdrawal group showed increases in caregiver-reported social motivation and improvements in emotion inference, alongside qualitative accounts of ``object-to-person'' reorientation (WDR3; TRF1--TRF3). In contrast, continued access was associated with more robot-centered social engagement and multiple accounts of siloed interaction during human-social opportunities (ENG2; TRF4--TRF5).

Importantly for child-centered ethics, withdrawal did not produce sustained distress in caregivers' accounts (ANX3; WDR2). This observation resonates with longitudinal discussions of how robots can persist as meaningful social artifacts and how their presence/absence can shape family narratives \cite{langer2023ethical}. In our context, the feasibility of withdrawal strengthens the argument for using it as a diagnostic evaluation tool in real-world settings: if the promise is scaffolding, the most revealing evidence should appear when the scaffold is removed.

\subsection{Mechanisms: the ``comfort trap'' and bounded learning in naturalistic home use}
\label{sec:discussion_mechanisms}
Our findings are consistent with (and we interpret through) a \emph{hypothesized} mechanism we term a \emph{comfort trap}: the same interaction properties that reduce anxiety (predictability, low-demand reciprocity, constant availability) may reduce the child's need to practice uncertainty tolerance and reciprocal negotiation in HHI. We do not directly measure uncertainty tolerance, caregiver co-use, or time allocation; thus, we treat this mechanism as an interpretive account rather than a causal claim. The continued-access group showed clear anxiolytic effects (ANX1--ANX2), consistent with prior work suggesting that structured, predictable robot interactions can support regulation and calm \cite{wu2025anxiety,matheus2022_ommie_deep_breathing,rakhymbayeva2021_long_term_engagement}. We emphasize that this anxiety reduction is a clinically meaningful benefit in its own right---particularly for autistic children, for whom co-occurring anxiety is a pervasive challenge; our argument is not that comfort is undesirable, but that affect regulation and transfer are separable goals that warrant simultaneous and independent design attention. Yet anxiety relief does not necessarily entail increased social seeking, and recent empirical work suggests that co-occurring anxiety and social motivation may relate in complex, non-linear ways for autistic children \cite{bagg2024socialmotivation}. In this light, affective benefits and transfer outcomes should be treated as separable design targets rather than assumed to align.

Qualitative accounts further suggest that emotion-related learning in robot contexts may become \emph{bounded} by the robot's representational format. Some caregivers described children who could succeed in robot-prompted labeling tasks but struggled to respond appropriately to real interpersonal distress (EMO3--EMO5). This pattern echoes concerns in the literature that in-context task performance can overestimate generalization when practice environments reduce variability and reciprocity \cite{kouroupa2022_socialrobots_meta,ghiglino2021followup}. In contrast, withdrawal appears to have increased exposure to everyday affect cues and the practical need for interpersonal negotiation, reflected in caregivers' reports of spontaneous comforting and increased sensitivity to tone and context (EMO1--EMO2).

At the ecological level, withdrawal may have perturbed household routines in ways that increased opportunities for HHI (e.g., caregivers compensating with shared activities; siblings becoming default partners). This can be interpreted in two ways: (1) as an alternative explanation (see Section~\ref{sec:discussion_alt}), and (2) as part of the phenomenon of interest. In home deployments, a robot is not merely a training tool; it is a \emph{participant} in an interaction ecology. The most design-relevant question becomes: \emph{what does the robot crowd out in the child's social ecology, and what does it catalyze?} \cite{david2025_lab_to_reality,matheus2025_long_term_interactions_review,langer2023ethical}. In our study, anxiety reductions under continued access co-occurred with weaker human-oriented trajectories relative to withdrawal, underscoring the need to design for handoff in addition to comfort.

\subsection{Implications for evaluation practice: measuring what the field claims to build}
\label{sec:discussion_evaluation}
A primary implication is evaluative: child-facing social agent research should avoid treating engagement, acceptability, or usability as proxies for developmental success. Recent systematic reviews have already emphasized heterogeneity in outcome measures and the difficulty of drawing strong conclusions about generalization \cite{kouroupa2022_socialrobots_meta,salimi2021_rcts_review,alabdulkareem2022_robot_assisted_therapy_review}. Building on this, we propose that evaluations should (a) pre-register \emph{transfer outcomes} as primary endpoints when scaffolding is the claim, and (b) explicitly incorporate \emph{ecological outcomes}---how time, attention, and social bids redistribute among caregivers, peers, and devices in the home.

Concretely, we recommend three evaluation practices for future work:
\begin{enumerate}[leftmargin=*,noitemsep]
    \item \textbf{Use withdrawal or interruption phases as transfer probes.} Studying the robot's absence can reveal whether observed gains persist and reappear in HHI contexts, and whether the system acted as a bridge or destination.
    \item \textbf{Pair engagement metrics with displacement-aware measures.} Alongside usability and enjoyment, measure changes in (i) human-directed social bids, (ii) caregiver co-use, and (iii) the allocation of interactive time across humans and devices.
    \item \textbf{Treat affect regulation and transfer as distinct outcomes.} Anxiety reduction is valuable, but should not be assumed to imply improved human social functioning. Evaluations should explicitly test both and report trade-offs when they arise \cite{wu2025anxiety,bagg2024socialmotivation}.
\end{enumerate}
These practices align with child-centered ethical guidance that emphasizes wellbeing, transparency, and accountability for AI systems interacting with children, especially in settings with reduced professional oversight \cite{unicef2025_ai_for_children_policy_guidance}.

\subsection{Design implications: ``transfer-first'' interaction patterns to preserve comfort without containment}
\label{sec:discussion_design}
Our results point toward a design goal that differs from the implicit ``best companion'' trajectory of many consumer social robots. For autism interventions, the goal should often be a \emph{scaffold designed to become less central over time}. Prior work has increasingly emphasized triadic family ecologies and caregiver-mediated approaches \cite{piccolo2024_parental_involvement_review,amirova2023_parental_involvement_ra_at,chen2025catalysts}. Our findings sharpen that agenda by highlighting substitution risk when the robot remains a highly attractive dyadic partner.

Based on the observed ``siloing'' versus ``handoff'' patterns, we propose the following transfer-first design patterns for home-deployed child social agents:
\begin{enumerate}[leftmargin=*,itemsep=2pt]
    \item \textbf{Triadic-by-default interaction.} Design activities that require a caregiver or sibling to complete (e.g., turn-taking that alternates between child and caregiver, or prompts that ask the child to query a human for input), shifting the unit of interaction from dyad to triad \cite{chen2025catalysts,piccolo2024_parental_involvement_review}.
    \item \textbf{Built-in handoffs (``show a human'' moments).} End each robot activity with a structured human-directed task (e.g., ``Tell your caregiver what you noticed about this face,'' ``Ask someone how they feel today''), creating repeated bridges from robot content to HHI.
    \item \textbf{Planned fading and off-ramps.} Introduce designed obsolescence: gradually reduce the robot's conversational dominance, shorten sessions, or add benign pauses that encourage the child to seek assistance or co-regulation from a human.
    \item \textbf{Anxiety relief that transitions to co-regulation.} Use the robot to initiate calming routines (breathing, grounding), but explicitly transition the routine to caregiver-led co-regulation after a short period \cite{matheus2022_ommie_deep_breathing}.
    \item \textbf{Transfer checks embedded in everyday life.} Provide lightweight caregiver prompts or weekly check-ins that ask whether a skill was used with a person that week. This creates accountability for transfer and reveals early signals of substitution risk.
\end{enumerate}
These patterns are not claims about Qrobot's internal design (which is commercial and not modifiable in our study); rather, they are design implications derived from a failure mode that can plausibly recur as consumer social agents become more capable, more available, and more companion-like.

\subsection{Alternative explanations and robustness considerations}
\label{sec:discussion_alt}
Several alternative explanations merit careful consideration. One interpretation is that the observed divergence reflects changes in \emph{human interaction opportunities} induced by the two conditions: continued access may have reduced the frequency or urgency of HHI opportunities, withdrawal may have increased opportunities through caregiver compensation, or both. In our study contacts, many withdrawal-group guardians described more frequent caregiver-led shared activities ($n = 11$), while others described substitutions to other screen media (tablet/TV; $n = 9$), underscoring that withdrawal can reorganize home routines in multiple directions (see Section~\ref{sec:procedure}). This could inflate apparent withdrawal benefits on caregiver-reported measures. However, we view this not merely as a nuisance variable but as part of the home ecology that designers must anticipate: consumer robots may reshape attention economies within families. Future work should measure caregiver time allocation and co-use explicitly, for example through short daily diaries or passive sensing of interaction episodes \cite{david2025_lab_to_reality,matheus2025_long_term_interactions_review}. More generally, we suggest treating ``attention economy'' variables (caregiver time, co-use frequency, and substitutions to other media/devices) as first-class outcomes and mediators in home deployments, rather than as post-hoc confounds. Doing so can separate robot-driven comfort benefits from caregiver-mediated compensation effects and potential displacement dynamics.

A related concern is expectation and reporting bias: because allocation could not be blinded, caregivers in the withdrawal condition were aware the robot had been removed and may have been more attuned to noticing or reporting social improvements, potentially inflating guardian-reported transfer outcomes. The two transfer outcomes most exposed to this bias are SMS and BES, which depend entirely on guardian perception. RMET---a child performance task scored by researchers---is substantially less susceptible to guardian-report bias; the fact that RMET divergence ($d = -1.08$) is directionally consistent with the guardian-reported SMS and BES provides important corroboration that the pattern is not solely an artifact of reporting bias. The concrete behavioral vignettes in qualitative accounts (e.g., spontaneous comforting episodes described with specific detail) further provide convergent evidence that is less abstracted than summary ratings and less amenable to post-hoc reinterpretation. Nonetheless, future studies should incorporate objective behavioral observations (e.g., coded parent--child interaction tasks, in-home video sampling with consent) to validate transfer mechanisms independent of guardian report. Additionally, we did not instrument Qrobot use with telemetry; therefore, dosage variation (time spent, whether co-used) could partially explain heterogeneity. This limitation is common in real-world deployments and strengthens the case for integrating usage logging or caregiver-reported dose measures into future trials \cite{matheus2025_long_term_interactions_review}.

Finally, the baseline table reveals a moderate imbalance in ASD history that warrants explicit interpretive caution: the withdrawal group had longer average ASD histories than the continued-access group (M = 20.80 vs.\ M = 15.15 months; SMD = 0.69, $p = .034$). This imbalance has substantive implications: children with longer intervention histories may have more developed social repertoires or greater readiness to reorient toward human interaction once the robot is absent, which could independently contribute to the withdrawal group's stronger transfer trajectories regardless of the withdrawal manipulation itself. While randomization reduces systematic bias on average, small samples can produce consequential imbalances. To assess robustness, we fit covariate-adjusted mixed-effects models including child age and ASD history months. Key Group$\times$Time interaction terms remained highly similar to the primary models and remained statistically significant across outcomes (Supplementary Tables~S13--S17), suggesting that the baseline imbalance alone is unlikely to fully account for the observed divergence. Nevertheless, this imbalance is a primary reason results should be interpreted cautiously, and future work should employ larger samples or pre-stratified randomization to ensure stronger baseline equivalence.

\subsection{Limitations and future directions}
\label{sec:discussion_limitations}
This study has limitations that should shape interpretation and future research. First, the sample size is modest, and participants were recruited through specific community channels, which may limit generalizability. Second, children had prior exposure to Qrobot not exceeding three months at enrollment. This criterion reduces pure novelty effects and ensures basic familiarity, but it may also over-represent families already willing and able to integrate the device into routines; effects may differ for first-time adopters and for longer-term use beyond three months. Third, our 8-week window captures near-term ecological shifts but cannot determine longer-term trajectories after withdrawal or extended exposure. Fourth, SMS, BES, and SCARED all rely on the same guardian as reporter, creating shared-method variance that may inflate within-group consistency across these three measures and potentially amplify between-group contrasts on guardian-reported outcomes. In the withdrawal condition specifically, guardians who knew the robot had been removed may have been more alert to changes in human-directed social behavior, increasing the risk of systematic over-reporting on SMS and BES relative to the continued-access group. RMET (child performance, researcher-scored) and concrete qualitative behavioral accounts partially mitigate this concern by providing independent evidence sources; future work should triangulate with structured observational data (e.g., coded parent--child interaction tasks) and/or sensor-derived behavioral measures. Fifth, the suitability of child self-report measures for younger children is a known challenge; age-restricted sensitivity analyses should be reported for RCADS. Sixth, RMET was administered at T1/T2/T3; repeated exposure can produce practice/retest effects and maturation-related gains, so absolute within-group change should be interpreted cautiously. Because both groups completed RMET on the same schedule, the increasing between-group divergence remains informative, but RMET should still be treated as a performance proxy rather than a direct observation of HHI behavior in naturalistic settings; future work should consider alternate forms (when available) and/or complement RMET with structured parent--child interaction tasks or in-home behavioral sampling. Finally, because Qrobot is a commercial product, we could not experimentally vary interaction design features to isolate which design properties most strongly drive anxiety relief versus transfer limitations. A next step is to test transfer-first design patterns in systems that support modifiable interaction policies and planned fade-out schedules, with longer follow-up and explicit measurement of displacement, caregiver mediation, and attention-economy variables \cite{piccolo2024_parental_involvement_review,chen2025catalysts,matheus2025_long_term_interactions_review}.

\section{Conclusion}
\label{sec:conclusion}
This work challenges a common assumption in child-facing HRI for autism: that sustained engagement with a social robot necessarily functions as a scaffold for later human--human interaction (HHI). In an 8-week home RCT of a commercial consumer robot, we observed a consistent divergence across outcome domains. Continued access was associated with affect-regulation benefits, reflected in lower guardian- and child-reported anxiety. At the same time, transfer-relevant outcomes favored withdrawal: social motivation and emotion inference improved more when the robot was absent, alongside qualitative evidence of increased human-directed social bids and more spontaneous real-world empathic responding. Crucially, these patterns occurred in the context of high perceived usability, underscoring a consequential failure mode: \emph{high engagement and smooth interaction are not sufficient evidence of transfer}. We contribute mixed-method evidence that engagement, affect regulation, and human-oriented transfer can move in different directions within the same deployment; a conceptual framing (\emph{Engagement is Not Transfer}) that distinguishes comfort and usability from transfer outcomes and ecological reallocation effects; and a methodological proposal to treat withdrawal as a \emph{transfer probe} that can reveal whether a robot operates as a bridge to HHI or as a potentially socially central destination. For design, our findings motivate \emph{transfer-first} interaction patterns---including triadic-by-default activities, explicit handoffs to caregivers and peers, and planned fading/off-ramps---that aim to preserve anxiolytic value without reinforcing robot-centered social containment.

More broadly, this study argues for a shift in how child-facing social agents are evaluated and designed for home life. The central question is not only whether children can engage with a robot, but how that engagement reshapes the relational ecosystem in which children develop. Designing for transfer requires making the bridge visible, measurable, and intentionally traversable.

\bibliographystyle{ACM-Reference-Format}
\bibliography{update0120}

\end{document}